\documentclass[twoside]{article}
\usepackage[accepted]{aistats2e}

\usepackage{microtype}
\usepackage{amsmath}
\usepackage{amssymb}
\usepackage[round]{natbib}
\newcommand{\citeg}[1]{\citep[e.g.][]{#1}}
\usepackage[small,tight]{bibhacks}
\usepackage{url}
\usepackage{color}
\usepackage{graphicx}
\usepackage{nicefrac}
\usepackage{booktabs}
\usepackage{mathclap}

\newcommand{\myvcenter}[1]{\ensuremath{\vcenter{\hbox{#1}}}} %

\newcommand{\gv}{\mathbf{f}} %
\newcommand{\gve}{f} %
\newcommand{\gvv}{\mathbf{g}} %
\newcommand{\bx}{\mathbf{x}}
\newcommand{\by}{\mathbf{y}}
\newcommand{\bmm}{\mathbf{m}}

\newcommand{\bnu}{\boldsymbol{\nu}}

\newcommand{\N}{\mathcal{N}}
\renewcommand{\O}{\mathcal{O}} %

\newcommand{\ave}[2]{{\mathbb E}_{#1}\kern-2pt\left[ #2 \right]}
\newcommand{\oo}[1]{\frac{1}{#1}} %

\newcommand{\te}{\!=\!} %
\newcommand{\tp}{\!+\!} %
\newcommand{\tm}{\!-\!} %

\newcommand{\tgt}{\!>\!}

\newcommand{\la}{\!\leftarrow\!}
\newcommand{\tsim}{\!\sim\!}

\newcommand{\g}{\!\mid\!} %

\newcommand{\thetamn}{\theta_\mathrm{min}}
\newcommand{\thetamx}{\theta_\mathrm{max}}

\newcommand{\posterior}{p^\star\kern-1pt} %

\begin{document}

\twocolumn[

\aistatstitle{Elliptical slice sampling}

\aistatsauthor{ Iain Murray \And Ryan Prescott Adams \And David J.C. MacKay}
\aistatsaddress{ University of Toronto \And University of Toronto \And University of Cambridge } ]

\runningtitle{Elliptical slice sampling}
\runningauthor{Murray, Adams, MacKay}

\begin{abstract}

Many probabilistic models introduce strong dependencies between
variables using a latent multivariate Gaussian distribution or a Gaussian
process. We present a new Markov chain Monte Carlo algorithm for performing inference in
models with multivariate Gaussian priors. Its key properties are: 1)~it has
simple, generic code applicable to many models, 2)~it has no free parameters,
3)~it works well for a variety of Gaussian process based models.
These properties make our method ideal for use while model building, removing the need
to spend time deriving and tuning updates for more complex algorithms.

\end{abstract}

\section{Introduction}

\vspace*{-0.01in}

The multivariate Gaussian distribution is commonly used to specify
a~priori beliefs about dependencies between latent variables in
probabilistic models.  The parameters of such a Gaussian may be
specified directly, as in graphical models and Markov random fields,
or implicitly as the marginals of a Gaussian process (GP)\@. Gaussian
processes may be used to express concepts of spatial or temporal
coherence, or may more generally be used to construct Bayesian
kernel methods for nonparametric regression and classification.
\Citet{rasmussen2005a} provide a recent review of GPs.%

Inferences can only be calculated in closed form for the simplest Gaussian
latent variable models. Recent work shows that posterior marginals can sometimes
be well approximated with deterministic methods \citep{kuss2005,rue2009}. Markov
chain Monte Carlo (MCMC) methods represent joint posterior distributions
with samples \citeg{neal1993}. MCMC can be slower but applies more
generally.

In some circumstances MCMC provides good results with minimal
model-specific implementation.  Gibbs sampling, in particular, is
frequently used to sample from probabilistic models in a
straightforward way, updating one variable at a time.  In
models with strong dependencies among variables, including many with Gaussian priors, Gibbs
sampling is known to perform poorly.  Several authors
have previously addressed the issue of sampling from models containing
strongly correlated Gaussians, notably the recent work of
\citet{titsias2009}.  In this paper we provide a technique called
\textit{elliptical slice sampling} that is simpler and often faster
than other methods, while also removing the need for preliminary
tuning runs.  Our method provides a drop-in replacement for MCMC
samplers of Gaussian models that are currently using Gibbs or
Metropolis--Hastings and we demonstrate empirical success against
competing methods with several different GP-based likelihood models.

\section{Elliptical slice sampling}
\vspace*{-0.05in}

Our objective is to sample from a
posterior distribution over latent variables that is proportional to
the product of a multivariate Gaussian prior and a likelihood function
that ties the latent variables to the observed data.  We will
use~$\gv$ to indicate the vector of latent variables that we wish to
sample and denote a zero-mean Gaussian distribution with
covariance~$\Sigma$~by\\[-0.1in]
\begin{equation}
    \N(\gv; 0,\Sigma) \equiv |2\pi\Sigma|^{-\nicefrac{1}{2}}\; \exp\kern-1pt\big(
    -\kern-2pt{\textstyle\frac{1}{2}}\, \gv^\top \Sigma^{-1} \gv
    \big).
    \label{eqn:gaussian}
\end{equation}\\[-0.15in]
We also use ${\gv \sim \N(0,\Sigma)}$ to state that $\gv$ is drawn from
a distribution with the density in~\eqref{eqn:gaussian}. Gaussians
with non-zero means can simply be shifted to have zero-mean with a
change of variables; an example will be given in
Section~\ref{sec:local}.  We use~${L(\gv)=p(\mathrm{data}\g\gv)}$ to
denote the likelihood function so that our \textit{target
distribution} for the MCMC sampler is\\[-0.1in]
\begin{equation}
    \posterior(\gv) = \frac{1}{Z}\; \N(\gv; 0,\Sigma)\,L(\gv),
    \label{eqn:posterior}
\end{equation}
where~$Z$ is the normalization constant, or the marginal likelihood, of
the model.

Our starting point is a Metropolis--Hastings method introduced by
\citet{neal1999a}. Given an initial state $\gv$, a new state
\begin{equation}
    \gv' = \sqrt{1-\epsilon^2}\; \gv + \epsilon\,\bnu, \qquad \bnu\sim\N(0,\Sigma)
    \label{eqn:underrelax}
\end{equation}
is proposed, where ${\epsilon\in[-1,1]}$ is a step-size parameter. The proposal is a
sample from the prior for $\epsilon\te1$ and more conservative for values closer
to zero. The move is accepted with probability
\begin{equation}
    p(\text{accept}) = \min(1,\,L(\gv')/L(\gv)),
    \label{eqn:mh_accept_prob}
\end{equation}
otherwise the next state in the chain is a copy of~%
$\gv$.

\citeauthor{neal1999a} reported that for some Gaussian process classifiers the
Metropolis--Hastings method was many times faster than Gibbs sampling. The
method is also simpler to implement and can immediately be applied to a much
wider variety of models with Gaussian priors.

A drawback, identified by \citet{neal1999a}, is that the
step-size~$\epsilon$ needs to be chosen appropriately for the Markov chain
to mix efficiently. This may require preliminary runs. Usually parameters
of the covariance $\Sigma$ and likelihood function $L$ are also inferred
from data.  Different step-size parameters may be needed as the model
parameters are updated. It would be desirable to automatically search over
the step-size parameter, while maintaining a valid algorithm.

For a fixed auxiliary random draw, $\bnu$,
the locus of possible proposals by varying
$\epsilon\in[-1,1]$ in \eqref{eqn:underrelax} is half of an ellipse.
\begin{equation}
    \gv' = \bnu\,\sin\theta + \gv\,\cos\theta,
    \label{eqn:ellipse}
\end{equation}
defining a full ellipse passing through the current state~$\gv$ and the
auxiliary draw~$\bnu$. For a fixed $\theta$ there is an equivalent $\epsilon$
that gives the same proposal distribution in the original algorithm. However, if
we can search over the step-size, the full ellipse gives a richer choice of
updates for a given~$\bnu$.

\subsection{Sampling an alternative model}
\label{sec:aux_model}

`Slice sampling' \citep{neal2003a} provides
a way to
sample along a line with an adaptive step-size. Proposals are drawn from an
interval or `bracket' which, if too large, is shrunk automatically until an
acceptable point is found. There are also
ways to automatically enlarge small initial brackets.
Naively applying these adaptive algorithms to
select the value of $\epsilon$ in~\eqref{eqn:underrelax} or $\theta$
in~\eqref{eqn:ellipse}
does not result in a Markov chain transition operator with the correct
stationary distribution. The locus of states is defined using the current
position $\gv$, which upsets the reversibility and correctness of the update.

We would like to construct a valid Markov chain transition operator on
the ellipse of states that uses slice sampling's existing ability to
adaptively pick step sizes. We will first intuitively construct a
valid method by positing an augmented probabilistic model in which the
step-size is a variable. Standard slice sampling algorithms then apply
to that model. We will then adjust the algorithm for our particular
setting to provide a second, slightly tidier algorithm.

Our augmented probabilistic model replaces the original latent
variable with prior $\gv\sim\N(0,\Sigma)$ with
\begin{equation}
\begin{split}
    \bnu_0 &\sim \N(0,\Sigma)\\
    \bnu_1 &\sim \N(0,\Sigma)\\
    \theta &\sim \mathrm{Uniform}[0,2\pi]\\
    \gv &= \bnu_0 \sin\theta + \bnu_1 \cos\theta.
\end{split}
\label{eqn:aux_model}
\end{equation}
The marginal distribution over the original latent variable $\gv$ is
still~$\N(0,\Sigma)$, so the distribution over data is identical. However, we
can now sample from the posterior over the new latent variables:
\begin{equation*}
    \newcommand{\xfn}{\gv\mbox{\small $(\bnu_0, \bnu_1, \theta)$}}
    \posterior(\bnu_0, \bnu_1, \theta) \propto \N(\bnu_0; 0,\Sigma)\, \N(\bnu_1; 0,\Sigma)\, L(\xfn),
\end{equation*}
and use the values of $\gv$ deterministically derived from these
samples. Our first approach applies two Monte Carlo transition
operators that leave the new latent posterior invariant.

\textbf{Operator~1:} jointly resample the latents $\bnu_0,\bnu_1,\theta$ given
the constraint that
$\gv(\bnu_0,\bnu_1,\theta)$
is unchanged. Because the
effective variable of interest doesn't change, the likelihood does not affect
this conditional distribution, so the update is generic and easy to implement.

\textbf{Operator~2:} use a standard slice sampling algorithm to update the
step-size $\theta$ given the other variables.

\begin{figure}
    \small

\vspace*{-0.09in}
\begin{tabular}{l}
\hphantom{\hspace*{0.9\linewidth}}\\
\toprule
\begin{minipage}{0.95\linewidth}
\textbf{Input:} current state $\gv$,
a routine that samples from $\N(0,\Sigma)$,
log-likelihood function $\log L$.

\smallskip

\textbf{Output:} a new state $\gv'$. When $\gv$ is drawn 
from $\posterior(\gv) \!\propto\! \N(\gv; 0,\Sigma)\,L(\gv)$, the marginal
distribution
of $\gv'$ is also~$\posterior$\rlap{.}
\end{minipage}\\
\midrule
\end{tabular}

\vspace*{-0.2cm}

\begin{enumerate}
        \small
\setlength{\itemsep}{1pt}
\setlength{\parskip}{0pt}
\setlength{\parsep}{0pt}
\item Sample from $p(\,\bnu_0,\bnu_1,\theta\g (\bnu_0\sin\theta \tp \bnu_1 \cos\theta \te \gv)\,)$:\\[-0.07in]
    \label{alg:alg1drawaux}
\begin{equation*}
\begin{split}
    \theta &\kern2pt\sim \mathrm{Uniform}[0,2\pi]\\[-0.02in]
      \bnu &\kern2pt\sim \N(0, \Sigma)\\[-0.02in]
    \bnu_0  &\leftarrow   \gv\sin\theta + \bnu\cos\theta\\[-0.02in]
    \bnu_1  &\leftarrow   \gv\cos\theta - \bnu\sin\theta\\[-0.02in]
\end{split}
\end{equation*}\\[-0.23in]
\item Update $\theta\!\in\![0,2\pi]$ using slice sampling \citep{neal2003a} on:\\[-0.09in]
\begin{equation*}
         \posterior(\theta \g \bnu_0,\bnu_1) \propto L(\bnu_0\sin\theta + \bnu_1 \cos\theta)
\end{equation*}\\[-0.3in]
\item \textbf{return} $\gv' = \bnu_0\sin\theta + \bnu_1 \cos\theta$
\end{enumerate}
\vspace*{-0.5cm}
\begin{tabular}{l}
\hphantom{\hspace*{0.95\linewidth}}\\
\bottomrule
\end{tabular}
\caption{\small Intuition behind elliptical slice sampling. This is a valid
algorithm, but will be adapted (Figure~\ref{alg:ess}).
}
\vspace*{-0.1in}
\label{alg:ess_intuition}
\end{figure}

The resulting algorithm
is given in Figure~\ref{alg:ess_intuition}. The auxiliary model
construction makes the link to slice sampling explicit, which makes it
easy to understand the validity of the approach. However, the
algorithm can be neater and the $[0,2\pi]$ range for slice sampling is
unnatural on an ellipse. The algorithm that we will present in detail
results from eliminating $\bnu_0$ and~$\bnu_1$ and a different way of
setting slice sampling's initial proposal range. The precise
connection will be given in Section~\ref{sec:generalizations}. A more
direct, technical proof that the equilibrium distribution of the
Markov chain is the target distribution is presented in
Section~\ref{sec:validity}.

\textbf{Elliptical slice sampling}, our proposed algorithm is given in
Figure~\ref{alg:ess}, which includes the details of the slice sampler.
An example run is illustrated in Figure~\hbox{\ref{fig:demo}(a--d)}.
Even for high-dimensional problems, the states considered within one
update lie in a two-dimensional plane. In high dimensions $\gv$
and~$\bnu$ are likely to have similar lengths and be an angle of
$\pi/2$ apart. Therefore the ellipse will typically be fairly close to
a circle, although this is not required for the validity of the
algorithm.

As intended, our slice sampling approach selects a new location on the randomly
generated ellipse in~\eqref{eqn:ellipse}. There are no rejections: the new state
$\gv'$ is never equal to the current state $\gv$ unless that is the only state
on the ellipse with non-zero likelihood. The algorithm proposes the angle
$\theta$ from a bracket $[\thetamn,\thetamx]$ which is shrunk exponentially
quickly until an acceptable state is found. Thus the step size is effectively
adapted on each iteration for the current $\bnu$ and~$\Sigma$.

\subsection{Computational cost}
\label{sec:costs}

Drawing $\bnu$ costs $\O(N^3)$, for $N$-dimensional $\gv$ and general~$\Sigma$.
The usual implementation of a Gaussian sampler would involve caching a
(Cholesky) decomposition of $\Sigma$, such that draws on subsequent iterations
cost $\O(N^2)$. For some problems with special structure drawing samples from
the Gaussian prior can be cheaper.

In many models the Gaussian prior distribution captures dependencies: the
observations are independent conditioned on~$\gv$. In these cases, computing
$L(\gv)$ will cost $\O(N)$ computation. As a result, drawing the $\bnu$ random
variate will be the dominant cost of the update in many high-dimensional problems.
In these cases the extra cost of elliptical slice sampling over Neal's
Metropolis--Hastings algorithm will be small.

As a minor performance improvement, our implementation optionally accepts the
log-likelihood of the initial state, if known from a previous update, so that it
doesn't need to be recomputed in step~\ref{alg:slice_height}.

\begin{figure}
    \small

\vspace*{-0.09in}
\begin{tabular}{l}
\hphantom{\hspace*{0.9\linewidth}}\\
\toprule
\begin{minipage}{0.95\linewidth}
\textbf{Input:} current state $\gv$,
a routine that samples from $\N(0,\Sigma)$,
log-likelihood function $\log L$.

\smallskip

\textbf{Output:} a new state $\gv'$. When $\gv$ is drawn 
from $\posterior(\gv) \!\propto\! \N(\gv; 0,\Sigma)\,L(\gv)$, the marginal
distribution
of $\gv'$ is also~$\posterior$\rlap{.}
\end{minipage}\\
\midrule
\end{tabular}

\vspace*{-0.2cm}

\begin{enumerate}
        \small
\setlength{\itemsep}{1pt}
\setlength{\parskip}{0pt}
\setlength{\parsep}{0pt}
\item Choose ellipse: $\bnu \kern2pt\sim \N(0, \Sigma)$ \label{alg:drawnu}
\item Log-likelihood threshold:\\[-0.07in]
\begin{equation*}
\begin{split}
         u  &\kern2pt\sim   \mathrm{Uniform}[0, 1]\\
    \log y  &\leftarrow   \log L(\gv) + \log u\\
\end{split}
\end{equation*}\\[-0.23in]
\label{alg:slice_height}
\item Draw an initial proposal, also defining a bracket:\\[-0.07in]
    \label{alg:initprop}
\begin{equation*}
\begin{split}
    \theta &\kern2pt\sim \mathrm{Uniform}[0,2\pi]\\
    [\thetamn,\thetamx] &\leftarrow [\theta \tm 2\pi,\, \theta]\\
\end{split}
\end{equation*}\\[-0.23in]
\item $\gv' \leftarrow \gv\cos\theta + \bnu \sin\theta$
      \label{alg_item:loop_point}
\item \textbf{if} $\log L(\gv') > \log y$ \textbf{then:}
\item \qquad Accept: \textbf{return} $\gv'$
\item \textbf{else:}
\item[] \qquad Shrink the bracket and try a new point:
\item \qquad\textbf{if} $\theta < 0$ \textbf{then:}
$\thetamn \la \theta$ \textbf{else:} $\thetamx \la \theta$
\label{alg:shrink}
\item \qquad$\theta \sim \mathrm{Uniform}[\thetamn,\thetamx]$
\item \qquad\textbf{GoTo} \ref{alg_item:loop_point}.
\end{enumerate}

\vspace*{-0.5cm}
\begin{tabular}{l}
\hphantom{\hspace*{0.95\linewidth}}\\
\bottomrule
\end{tabular}

\caption{\small The elliptical slice sampling algorithm.
}
\label{alg:ess}
\end{figure}
\begin{figure}
    \vspace*{0.4cm}

    \hspace*{0.5cm}\hbox{\hbox{\includegraphics[scale=0.65]{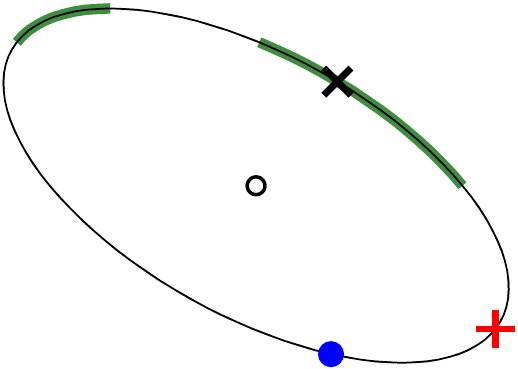}\llap{\raisebox{0.6cm}{(a)}\hspace*{3.2cm}}}\hspace*{0.5cm}
    \hbox{\includegraphics[scale=0.65]{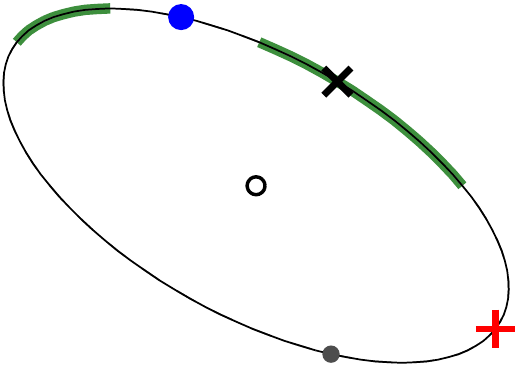}\llap{\raisebox{0.6cm}{(b)}\hspace*{3.2cm}}}}

    \vspace*{0.3cm}

    \hspace*{0.5cm}\hbox{\hbox{\includegraphics[scale=0.65]{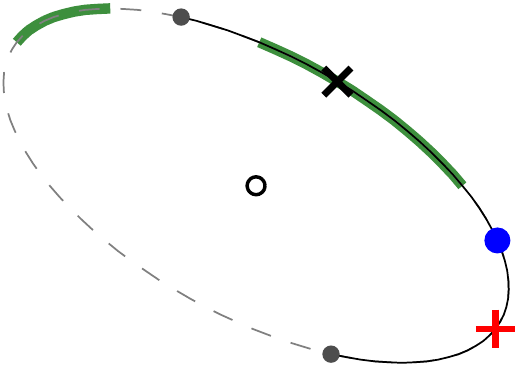}\llap{\raisebox{0.6cm}{(c)}\hspace*{3.2cm}}}\hspace*{0.5cm}
    \hbox{\includegraphics[scale=0.65]{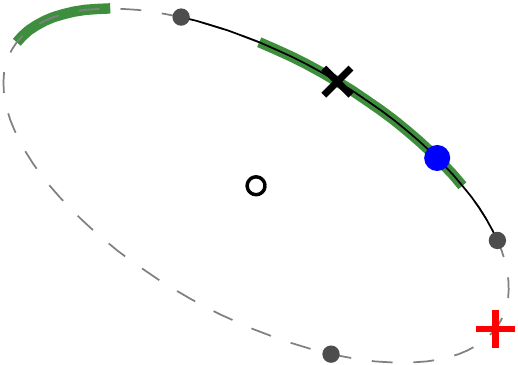}\llap{\raisebox{0.6cm}{(d)}\hspace*{3.2cm}}}}

    \vspace*{0.3cm}

    \hspace*\fill
    \hbox{\includegraphics[scale=0.65]{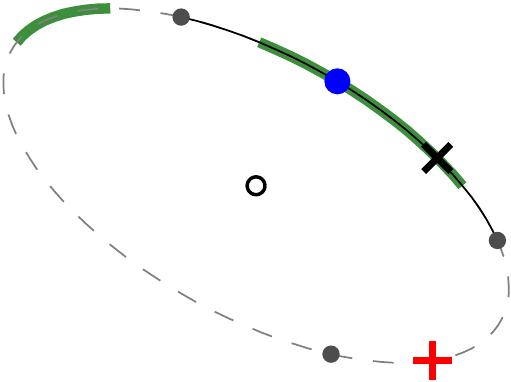}\llap{\raisebox{0.6cm}{(e)}\hspace*{3.2cm}}}\hspace*{0.25cm}

    \vspace*{-0.3cm}

    \caption{\small \textbf{(a)}~The algorithm receives
    $\gv${$=$}\includegraphics[scale=0.65]{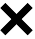} as input.
    Step~\ref{alg:drawnu} draws auxiliary variate
    $\bnu${$=$}\smash{\includegraphics[scale=0.65]{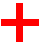}}, defining an
    ellipse centred at the
    origin~(\includegraphics[scale=0.65]{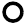}).
    Step~\ref{alg:slice_height}: a likelihood threshold defines the
    `slice'~(\myvcenter{\includegraphics[height=0.8ex,width=1.3em]{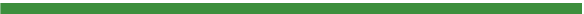}}).
    Step~\ref{alg:initprop}: an initial
    proposal~\includegraphics[scale=0.65]{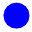} is drawn, in
    this case not on the slice. \textbf{(b)}~The first proposal defined
    both edges of the $[\thetamn,\thetamx]$ bracket; the second
    proposal~(\includegraphics[scale=0.65]{ellipses/key_cur_prop}) is also drawn
    from the whole range. \textbf{(c)}~One edge of the
    bracket~(\myvcenter{\includegraphics[height=0.4ex,width=1.3em]{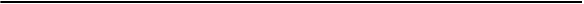}})
    is moved to the last rejected point such that
    \includegraphics[scale=0.65]{ellipses/key_x} is still included. Proposals
    are made with this shrinking rule until one lands on the slice.
    \textbf{(d)}~The proposal
    here~(\includegraphics[scale=0.65]{ellipses/key_cur_prop}) is on the slice
    and is returned as~$\gv'$. \textbf{(e)}~Shows the reverse configuration
    discussed in Section~\ref{sec:validity}:
    \includegraphics[scale=0.65]{ellipses/key_x} is the input $\gv'$, which with
    auxiliary $\bnu'${$=$}\smash{\includegraphics[scale=0.65]{ellipses/key_nu}}
    defines the same ellipse. The brackets and first three
    proposals~(\includegraphics[scale=0.65]{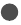}) are the
    same. The final
    proposal~(\includegraphics[scale=0.65]{ellipses/key_cur_prop}) is accepted,
    a move back to~$\gv$.
    } \label{fig:demo}
\end{figure}
\subsection{Validity}%
\label{sec:validity}

Elliptical slice sampling considers settings of an angle variable,~$\theta$.
Figure~\ref{alg:ess} presented the algorithm as it would be used: there is no
need to index or remember the visited angles. For the purposes of analysis we
will denote the ordered sequence of angles considered during the algorithm
by~$\{\theta_k\}$ with $k\te1..K$.

We first identify the joint distribution over a state drawn from the target
distribution~\eqref{eqn:posterior} and the other random quantities generated by
the algorithm:
\begin{equation*}
    p(\gv,y,\bnu,\{\theta_k\})
    = \posterior(\gv)\,p(y\g\gv)\,p(\bnu)\,p(\{\theta_k\} \g \gv,\bnu,y)
\end{equation*}
\begin{equation}
    = \frac{1}{Z}\,\N(\gv;0,\Sigma)\,\N(\bnu;0,\Sigma)\,p(\{\theta_k\} \g \gv,\bnu,y),
    \label{eqn:forwards}
\end{equation}
where the vertical level $y$ was drawn uniformly in~$[0,L(\gv)]$, that is,
$p(y\g\gv)\te1/L(\gv)$. The final term, $p(\{\theta_k\} \g \gv,\bnu,y)$, is a
distribution over a random-sized set of angles, defined by the stopping rule
of the algorithm.

Given the random variables in \eqref{eqn:forwards} the algorithm
deterministically computes positions, $\{\gv_k\}$, accepting the first one that
satisfies a likelihood constraint. More generally each angle specifies a
rotation of the two a~priori Gaussian variables:
\begin{equation}
    \begin{split}
    \bnu_k &= \bnu\cos\theta_k -\gv\sin\theta_k \\
    \gv_k &= \bnu\sin\theta_k + \gv\cos\theta_k
    ,\quad k=1..K.
    \end{split}
    \label{eqn:rotation}
\end{equation}
For any choice of $\theta_k$ this deterministic transformation has unit Jacobian.
Any such rotation also leaves the joint prior probability invariant,
\begin{equation}
    \N(\bnu_k;0,\Sigma)\,
    \N(\gv_k;0,\Sigma) =
    \N(\bnu;0,\Sigma)\,
    \N(\gv;0,\Sigma)
    \label{eqn:prior_invariance}
\end{equation}
for all $k$, which can easily be verified by substituting values into the Gaussian
form~\eqref{eqn:gaussian}.

It is often useful to consider how an MCMC algorithm could make a \emph{reverse
transition} from the final state $\gv'$ back to the initial state~$\gv$. The
final state $\gv'\te\gv_K$ was the result of a rotation by $\theta_K$
in~\eqref{eqn:rotation}. Given an initial state of $\gv'\te\gv_K$, the algorithm
could generate~${\bnu'\te\bnu_K}$ in step~\ref{alg:drawnu}. Then a rotation of
$-\theta_K$ would return back to the original $(\gv,\bnu)$ pair. Moreover, the
same ellipse of states is accessible and rotations of $\theta_k\tm\theta_K$ will
reproduce any intermediate $\gv_{k<K}$ locations visited by the initial run of
the algorithm.

In fact, the algorithm %
is \emph{reversible}:
\begin{equation}
    p(\gv,y,\bnu,\{\theta_k\}) = p(\gv',y,\bnu',\{\theta_k'\}),
    \label{eqn:reversibility}
\end{equation}
the equilibrium probability of a forwards draw~\eqref{eqn:forwards} is the same
as the probability of starting at $\gv'$, drawing the same~$y$ (possible because
$L(\gv')\tgt y$),~${\bnu'\te\bnu_K}$ and
\begin{equation}
    \text{angles,}~\theta_k' = \begin{cases}
        \theta_k - \theta_K & k < K\\
        -\theta_K & k = K,
    \end{cases}
    \label{eqn:theta_shifts}
\end{equation}
resulting in the original state $\gv$ being returned. The reverse configuration
corresponding to the result of a forwards run in Figure~\hbox{\ref{fig:demo}(d)}
is illustrated in Figure~\hbox{\ref{fig:demo}(e)}. Substituting
\eqref{eqn:prior_invariance} into \eqref{eqn:forwards} shows that ensuring that
the forward and reverse angles are equally probable,
\begin{equation}
    p(\{\theta_k\} \g \gv,\bnu,y) = p(\{\theta_k'\} \g \gv',\bnu',y),
    \label{eqn:angles_same}
\end{equation}
results in the reversible property~\eqref{eqn:reversibility}.

The algorithm does satisfy \eqref{eqn:angles_same}:
The probability of the first angle is always $\nicefrac{1}{2\pi}$. If more
angles were considered before an acceptable state was found, these angles were
drawn with probabilities $1/(\thetamx-\thetamn)$. Whenever the bracket was
shrunk in step~\ref{alg:shrink}, the side to shrink must have been chosen such
that $\gv_K$ remained selectable as it was selected later. The reverse
transition uses the same intermediate proposals, making the same rejections
with the same likelihood threshold,~$y$. Because the algorithm
explicitly includes the initial state, which in reverse is $\gv_K$ at
$\theta'\te0$, the reverse transition involves the same set of shrinking
decisions as the forwards transitions. As the same brackets are sampled,
the $1/(\thetamx-\thetamn)$ probabilities for drawing angles are the same for
the forwards and reverse transitions.

The reversibility of the transition operator \eqref{eqn:reversibility} implies
that the target posterior distribution \eqref{eqn:posterior} is a stationary
distribution of the Markov chain. Drawing $\gv$ from the stationary distribution
and running the algorithm draws a sample from the joint auxiliary
distribution~\eqref{eqn:forwards}. The deterministic transformations in
\eqref{eqn:rotation} and~\eqref{eqn:theta_shifts} have unit Jacobian, so the
probability density of obtaining a joint draw corresponding
to~${(\gv',y,\bnu',\{\theta_k'\})}$
is equal to the probability given by
\eqref{eqn:forwards} for the original variables. The reversible property in
\eqref{eqn:reversibility} shows that this is the same probability as generating
the variables by first generating $\gv'$ from the target distribution and
generating the remaining quantities using the algorithm. Therefore, the marginal
probability of $\gv'$ is given by the target posterior~\eqref{eqn:posterior}.

Given the first angle, the distribution over the first proposed move is
$\N(\gv\cos\theta,\, \Sigma\sin^2\theta)$. Therefore, there is a non-zero
probability of transitioning to any region that has non-zero probability under
the posterior. This is enough to ensure that, formally, the chain is irreducible
and aperiodic \citep{tierney1994a}. Therefore, the Markov chain has a unique stationary distribution
and repeated applications of
elliptical slice sampling to an arbitrary starting point will asymptotically
lead to points drawn from the target posterior distribution~\eqref{eqn:posterior}.

\subsection{Slice sampling variants}
\label{sec:generalizations}

There is some amount of choice in how the slice sampler on the ellipse could be
set up. Other methods for proposing angles could have been used, as long as they
satisfied the reversible condition in~$\eqref{eqn:angles_same}$. The particular
algorithm proposed in Figure~\ref{alg:ess} is appealing because it is simple and
has no free parameters.

The algorithm must choose the initial edges of the bracket $[\thetamn,\thetamx]$
randomly. It would be aesthetically pleasing to place the edges of the bracket
at the opposite side of the ellipse to the current position, at $\pm\pi$.
However this deterministic bracket placement would not be reversible and gives
an invalid algorithm.

The edge of a randomly-chosen bracket could lie on the `slice', the
acceptable region of states. Our recommended elliptical slice sampling
algorithm, Figure~\ref{alg:ess}, would accept this point. The
initially-presented algorithm, Figure~\ref{alg:ess_intuition},
effectively randomly places the endpoints of the bracket but without
checking this location for acceptability. Apart from this small
change, it can be shown that the algorithms are equivalent.

In typical problems the slice will not cover the whole ellipse. For example, if
$\gv$ is a representative sample from a posterior, often $-\gv$ will not be.
Increasing the probability of proposing points close to the current state may
increase efficiency. One way to do this would be to shrink the bracket more
aggressively \citep{skilling2003}. Another would be to derive a model from the
auxiliary variable model~\eqref{eqn:aux_model}, but with a non-uniform
distribution on~$\theta$. Another way would be to randomly position an initial
bracket of width less than $2\pi$\,---\,the code that we provide optionally
allows this. However, as explained in section~\ref{sec:costs}, for
high-dimensional problems such tuning will often only give small improvements.
For smaller problems we have seen it possible to improve the cpu-time efficiency
of the algorithm by around two times.

Another possible line of research is methods for biasing proposals away from the
current state. For example the `over-relaxed' methods discussed by
\citet{neal2003a} have a bias towards the opposite side of the slice from the
current position. In our context it may be desirable to encourage moves close to
$\theta\te\pi/2$, as these moves are independent of the previous position. These
proposals are only likely to be useful when the likelihood terms are very
weak, however. In the limit of sampling from the prior due to a constant
likelihood, the algorithm already samples reasonably efficiently. To see this,
consider the distribution over the outcome after $N$~iterations initialized at
$\gv^{0}$:
\begin{equation*}
    \gv^{N} = \gv^{0}\prod_{n=1}^N \cos\theta^{n} + \sum_{m=1}^N
    \bnu^{m} \sin\theta^{m}\,\prod_{\mathclap{n=m+1}}^N \,\cos\theta^{n},
\end{equation*}
where $\bnu^n$ and $\theta^{n}$ are values drawn at iteration~$n$. Only one
angle is drawn per iteration when sampling from the prior, because the first
proposal is always accepted. The only dependence on the initial state is the
first term, the coefficient of which shrinks towards zero exponentially quickly.

\subsection{Limitations}

A common modeling situation is that an unknown constant offset,
${c\sim\N(0,\sigma^2_m)}$, has been added to the entire
latent vector~$\gv$. The resulting variable, $\gvv\te\gv\tp c$, is still
Gaussian distributed, with the constant $\sigma^2_m$ added to every element of
the covariance matrix. \citet{neal1999a} identified that this sort of covariance
will not tend to produce useful auxiliary draws~$\bnu$. An iteration of the
Markov chain can only add a nearly-constant shift to the current state. Indeed,
covariances with large constant terms are generally problematic as
they tend to be poorly conditioned. Instead, large offsets should be modeled and
sampled as separate variables.

No algorithm can sample effectively from arbitrary distributions. As any
distribution can be factored as in \eqref{eqn:posterior}, there exist
likelihoods~$L(\gv)$ for which elliptical slice sampling is not effective. Many
Gaussian process applications have strong prior smoothness constraints and
relatively weak likelihood constraints. This important regime is where we focus
our empirical comparison.

\section{Related work}
\label{sec:related}

Elliptical slice sampling builds on a Metropolis--Hastings (M--H) update
proposed by \citet{neal1999a}. Neal reported that the original update performed
moderately better than using a more obvious M--H proposal,
\begin{equation}
    \gv' = \gv + \epsilon\,\bnu, \qquad \bnu\sim\N(0,\Sigma),
    \label{eqn:mh1}
\end{equation}
and much better than Gibbs sampling for Gaussian process classification. Neal
also proposed using Hybrid/Hamiltonian Monte Carlo \citep{duane1987,neal1993},
which can be very effective, but requires tuning and the implementation of
gradients. We now consider some other alternatives that have similar
requirements to elliptical slice sampling.

\subsection{`Conventional' slice sampling}

Elliptical slice sampling builds on the family of methods introduced by
\citet{neal2003a}. Several of the existing slice
sampling methods would also be easy to apply: they only require point-wise evaluation
of the posterior up to a constant. These methods do have step-size parameters, but
unlike simple Metropolis methods, typically the performance of slice samplers
does not crucially rely on carefully setting free parameters.

The most popular generic slice samplers use simple univariate updates, although
applying these directly to~$\gv$ would suffer the same slow convergence problems
as Gibbs sampling. While \citet{agarwal2005} have applied slice sampling for
sampling parameters in Gaussian spatial process models, they assumed a
linear-Gaussian observation model. For non-Gaussian data it was suggested that
``there seems to be little role for slice sampling.''

Elliptical slice sampling changes all of the variables in $\gv$ at once,
although there are potentially better ways of achieving this. An extensive
search space of possibilities includes the suggestions for multivariate updates
made by \citet{neal2003a}.

One simple possible slice sampling update performs a univariate update
along a random line traced out by varying $\epsilon$
in~\eqref{eqn:mh1}. As the M--H method based on the line worked less
well than that based on an ellipse, one might also expect a line-based
slice sampler to perform less well. Intuitively, in high dimensions
much of the mass of a Gaussian distribution is in a thin ellipsoidal
shell. A straight line will more rapidly escape this shell than an
ellipse passing through two points within it.

\subsection{Control variables}
\label{sec:control}

\citet{titsias2009} introduced a sampling method inspired by sparse Gaussian
process approximations.  $M$~control variables $\gv_c$ are introduced such that
the joint prior $p(\gv,\gv_c)$ is Gaussian, and that~$\gv$
still has marginal prior~$\N(0,\Sigma)$. For Gaussian process models a
parametric family of joint covariances was defined, and the model is optimized
so that the control variables are informative about the original
variables:~$p(\gv\g\gv_c)$ is made to be highly peaked. The optimization is a
pre-processing step that occurs before sampling begins.

The idea is that the small number of control variables~$\gv_c$ will be less
strongly coupled than the original variables, and so can be moved individually
more easily than the components of~$\gv$. A proposal involves resampling one
control variable from the conditional prior and then resampling $\gv$ from
$p(\gv\g\gv_c)$. This move is accepted or rejected with the
Metropolis--Hastings rule.

Although the method is inspired by an approximation used for large datasets, the
accept/reject
step uses
the full model. After $\O(N^3)$
pre-processing it costs $\O(N^2)$ to evaluate a proposed change to the
$N$-dimensional vector~$\gv$. One `iteration' in the paper consisted of an
update for each control variable and so costs~$\O(MN^2)$ --- roughly $M$
elliptical slice sampling updates. The control method uses fewer likelihood
evaluations per iteration, although has some different minor costs associated
with book-keeping of the control variables.

\subsection{Local updates}
\label{sec:local}

In some applications it may make sense to update only a subset of the latent
variables at a time. This might help for computational reasons given the
$\O(N^2)$ scaling for drawing samples of subsets of size~$N$.
\citet{titsias2009} also identified suitable subsets for local updates and then
investigated sampling proposals from the conditional Gaussian prior.

In fact, local updates can be combined with any transition operator for models
with Gaussian priors. If $\gv_A$ is a subset of variables to update and $\gv_B$
are the remaining variables, we can write the prior as:
\begin{equation}
    \begin{bmatrix}
        \gv_A\\
        \gv_B
    \end{bmatrix} \sim
    \N\left( 0,
    \begin{bmatrix}
        \Sigma_{A,A} & \Sigma_{A,B}\\
        \Sigma_{B,A} & \Sigma_{B,B}
    \end{bmatrix}
    \right)
\end{equation}
and the conditional prior is:
\begin{equation*}
\begin{split}
    &p(\gv_A \g \gv_B) =\N(\gv_A;\, \bmm, S), \text{~where}\\
    &\bmm = \Sigma_{A,B} \Sigma_{B,B}^{-1}\gv_B,\text{~and~}
    S = \Sigma_{A,A} \tm \Sigma_{A,B} \Sigma_{B,B}^{-1} \Sigma_{B,A}.
\end{split}
\end{equation*}
A change of variables $\gvv\te\gv_A\tm\bmm$ allows
us to express the conditional posterior as:
    $\posterior(\gvv) \propto \N\kern-2pt\left(\gvv;\,0,\,S\right)
    L\big(
    \mbox{\tiny $\begin{bmatrix}
        \gvv \tp \bmm\\
        \gv_B
    \end{bmatrix}$}
    \big)$.
We can then apply elliptical slice sampling, or any alternative, to update
$\gvv$ (and thus $\gv_A$). Updating groups of variables according to their
conditional distributions is a standard way of sampling from a joint
distribution.

\section{Experiments}

We performed an empirical comparison on three Gaussian process based
probabilistic modeling tasks. Only a brief description of the models and methods
can be given here. Full code to reproduce the results
is provided as supplementary material.

\subsection{Models}

Each of the models associates a dimension of the latent variable, $\gve_n$, with
an `input' or `feature' vector~$\bx_n$. The models in our experiments construct
the covariance from the inputs using the most common method,
\begin{equation}
    \Sigma_{ij}
    \,=\, \sigma_f^2\kern1pt \exp\!\big(\kern-1pt-\kern-1pt{\textstyle\oo{2}
    \sum_{d=1}^D} (x_{d,i} - x_{d,j})^2 / \ell^2
    \,\big),
    \label{eqn:se_kernel}
\end{equation}
the squared-exponential or ``Gaussian'' covariance. This covariance has
``lengthscale'' parameter~$\ell$ and an overall ``signal
variance''~$\sigma_f^2$. Other covariances may be more appropriate in many
modeling situations, but our algorithm would apply unchanged.

\textbf{Gaussian regression:} given observations $\by$ of the latent variables
with Gaussian noise of variance $\sigma_n^2$,
\begin{equation}
    L_r(\gv) = {\textstyle\prod_n}\; \N(y_n; \gve_n, \sigma_n^2),
\end{equation}
the posterior is Gaussian and so fully tractable. We use this as a simple test
that the method is working correctly. Differences in performance on this task
will also give some indication of performance with a simple log-concave
likelihood function.

We generated ten synthetic datasets with input feature dimensions from one to
ten. Each dataset was of size~${N\te200}$, with inputs $\{\bx_n\}_{n=1}^N$ drawn
uniformly from a $D$-dimensional unit hypercube and function values drawn from a
Gaussian prior, $\gv\tsim\N(0,\Sigma)$, using covariance \eqref{eqn:se_kernel}
with lengthscale $\ell\te1$ and unit signal variance,~$\sigma^2_f\te1$.
Noise with variance $\sigma_n^2\te0.3^2$ was added to generate the observations.

\textbf{Gaussian process classification:} a well-explored application of
Gaussian processes with a non-Gaussian noise model is binary classification:
\begin{equation}
    L_c(\gv) = {\textstyle\prod_n}\; \sigma\left( y_n \gve_n \right),
\end{equation}
where $y_n\in\{-1,+1\}$ are the label data and $\sigma(a)$ is a sigmoidal
function: $1/(1+e^{-a})$ for the logistic classifier; a cumulative Gaussian for
the probit classifier.

We ran tests on the USPS classification problem as set up by
\citet{kuss2005}.
We used~${\log\sigma_f\te 3.5}$, ${\log\ell\te 2.5}$ and the logistic likelihood.

\textbf{Log Gaussian Cox process:} an inhomogeneous Poisson process with a
non-parametric rate can be constructed by using a shifted draw from a Gaussian
process as the log-intensity function. Approximate inference can be performed by
discretizing the space into bins and assuming that the log-intensity is uniform
in each bin \citep{moller1998}. Each bin contributes a Poisson
likelihood:
\begin{equation}
    L_p(\gv) = \prod_n \frac{{\lambda_n}^{y_n}\exp(-\lambda_n)}{y_n!}
    ,\quad \lambda_n \te e^{\gve_n + m},
    \label{eqn:plikelihood}
    \vspace*{-0.05in}
\end{equation}
where the model explains the $y_n$ counts in bin~$n$ as drawn from a Poisson
distribution with mean~$\lambda_n$. The offset to the log mean, $m$, is the mean
log-intensity of the Poisson process plus the log of the bin size.

We perform inference for a Cox process model of the dates of mining
disasters taken from a standard data set for testing point processes
\citep{jarrett1979}. The 191~events were placed into 811~bins of 50~days
each. The Gaussian process parameters were fixed to $\sigma^2_f\te1$ and
$\ell\te13516$ days (a third of the range of the dataset). The offset~$m$ in
\eqref{eqn:plikelihood} was set to~${m\te\log(191/811)}$, to match the empirical
mean~rate.

\subsection{Results}

\begin{figure}
    \hspace*\fill\includegraphics[width=0.9\linewidth]{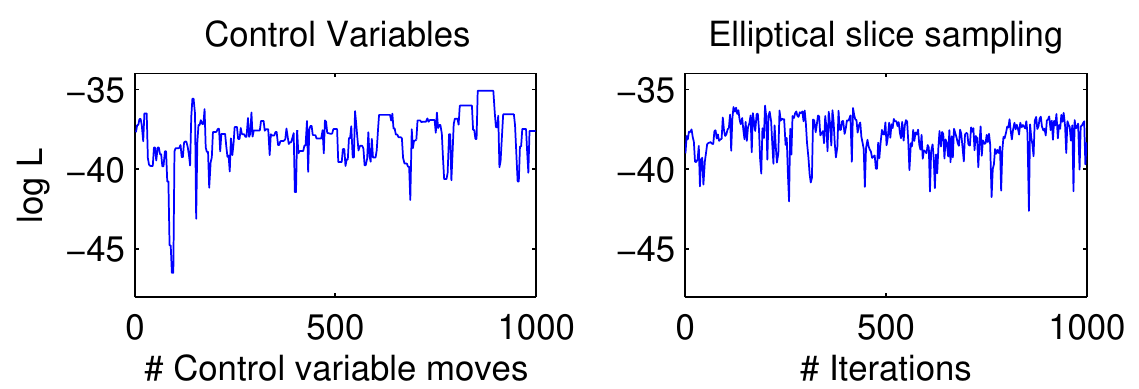}\hspace*\fill\\[-0.2in]
    \caption{\small Traces of log-likelihoods for the 1-dimensional GP
    regression experiment. Both lines are made with 333 points plotted after
    each sweep through $M\te3$ control variables and after every 3 iterations of
    elliptical slice sampling.}
    \label{fig:traces}
    \vspace*{-0.1in}
\end{figure}
\begin{figure}
    \hspace*\fill\includegraphics[width=\linewidth]{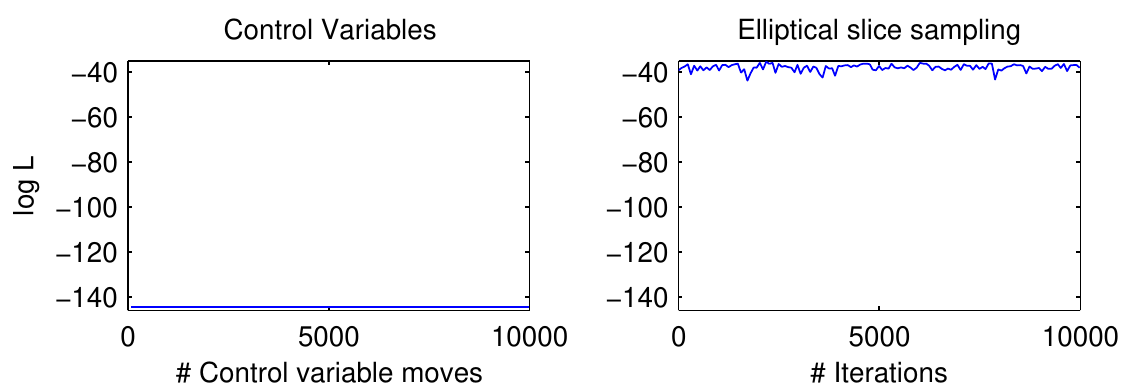}\hspace*\fill\\[-0.2in]
    \caption{\small As in Figure~\ref{fig:traces} but for 10-dimensional
    regression and plotting every $M\te78$ iterations. (Control variables didn't
    move on this run.)}
    \label{fig:traces10d}
    \vspace*{-0.1in}
\end{figure}

A trace of the samples' log-likelihoods, Figure~\ref{fig:traces}, shows that
elliptical slice sampling and control variables sampling have different
behavior. The methods make different types of moves and only control variables
sampling contains rejections. Using long runs of either method to estimate
expectations under the target distribution is valid. However, sticking in a
state due to many rejections can give a poor estimator as can always making
small moves. It can be difficult to judge overall sampling quality from trace
plots alone.

\begin{figure*}[t!]
  \centering%
  \includegraphics[width=\textwidth]{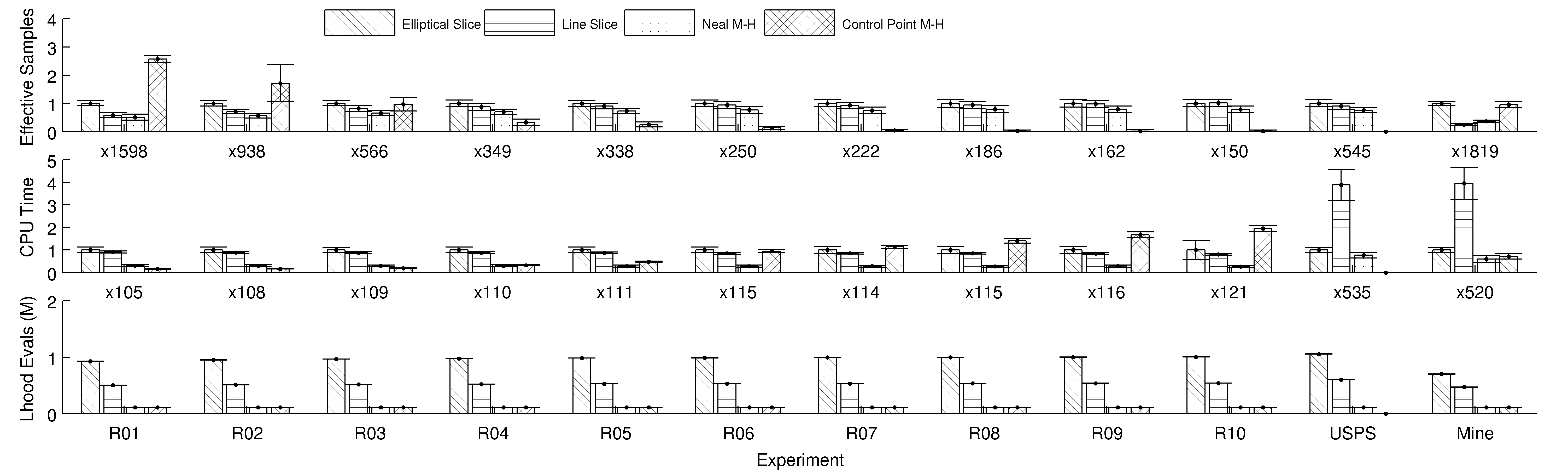}\\[-0.12in]
  \caption{\small Number of effective samples from $10^5$~iterations after
  $10^4$ burn in, with time and likelihood evaluations required. The means and
  standard deviations for 100 runs are shown (divide the ``error bars'' by 10 to
  get standard errors on the mean, which are small). Each iteration involves one
  $\O(N^2)$ operation (e.g.\ one $\bnu$ draw or updating one control variable).
  Each group of bars in the top two rows has been rescaled for readability: the
  numbers beneath each group show the number of effective samples or CPU
  time in seconds for elliptical slice sampling, which always has bars of
  height~1.}%
  \label{fig:bars}%
  \vspace*{-0.09in}
\end{figure*}

As a quantitative measure of quality
we estimated the ``effective number of samples'' from log-likelihood traces
using R-CODA \citep{cowles2006}. Figure~\ref{fig:bars} shows these results along
with computer time taken. The step size for Neal's Metropolis method was chosen
using a grid search to maximize performance. Times are for the provided
implementations under Matlab~v7.8 on a single $64\,$bit, $3\,$GHz Intel
Xeon~CPU\@. Comparing runtimes is always problematic, due to
implementation-specific details. Our numbers of effective samples are primarily
plotted for the same number of $\O(N^2)$ updates with the understanding that
some correction based loosely on runtime should be applied.

The control variables approach was particularly recommended for Gaussian
processes with low-dimensional input spaces. On our particular low-dimensional
synthetic regression problems using control variables clearly outperforms all
the other methods. On the model of mining disasters, control variable sampling
has comparable performance to elliptical slice sampling with about 50\% less run
time. On higher-dimensional problems more control variables are required; then
other methods cost less. Control variables failed to sample in
high-dimensions (Figure~\ref{fig:traces10d}). On the USPS classification problem
control variables ran exceedingly slowly and we were unable to obtain any
meaningful results.\looseness=-1

Elliptical slice sampling obtained more effective samples than Neal's M--H
method \emph{with the best possible step size}, although at the cost of
increased run time. On the problems involving real data, elliptical slice
sampling was better overall whereas M--H has more effective samples per unit
time (in our implementation) on the synthetic problems. The performance
differences aren't huge; either method would work well enough.

Elliptical slice sampling takes less time than slice sampling along a straight
line (line sampling involves additional prior evaluations) and usually
performs better.

\section{Discussion}

The slice samplers use many more likelihood evaluations than the other methods.
This is partly by choice: our code can take a step-size parameter to reduce the
number of likelihood evaluations (Section~\ref{sec:generalizations}). On these
problems the time for likelihood computations isn't completely negligible:
speedups of around $\times2$ may be possible by tuning elliptical slice
sampling. Our default position is that ease-of-use and human time is important
and that the advantage of having no free parameters should often be taken in
exchange for a factor of two in runtime.

We fixed the parameters of $\Sigma$ and~$L$ in our experiments to simplify the
comparison. Fixing the model potentially favors the methods that have adjustable
parameters.
In problems were $\Sigma$ and~$L$ change
dramatically, a single step-size or optimized set of control variables could
work very poorly.

Elliptical slice sampling is a simple generic algorithm with no tweak
parameters.
It performs similarly to the best possible performance of a related M--H scheme,
and could be applied to a wide variety of applications in both low and high
dimensions.

\subsubsection*{Acknowledgements}

\vspace*{-0.07in}

Thanks to Michalis Titsias for code, Sinead Williamson and Katherine Heller for
a helpful discussion, and to Radford Neal, Sam Roweis, Christophe Andrieu and
the reviewers for useful suggestions. RPA is funded by the Canadian Institute
for Advanced Research.

\vspace*{-0.05in}

\let\origurl\url
\renewcommand{\url}[1]{\penalty10000 \hskip.5em
    plus\linewidth \interlinepenalty10000\penalty200
    \hskip-.17em plus-\linewidth minus.11em \origurl{#1}}

\bibliographystyle{abbrvnat} %
\bibliography{arxiv}

\end{document}